\documentclass[prb,twocolumn,showpacs,superscriptaddress,groupedaddress]{revtex4-1}

\usepackage {graphicx}
\usepackage {amsmath}
\usepackage {amsfonts}
\usepackage {amssymb}
\usepackage {mathrsfs}
\usepackage {bbm}
\usepackage{subfigure}

\newcommand {\be}{\begin{eqnarray}}
\newcommand {\ee}{\end{eqnarray}}

\begin{document}

\title {Three-band superconductivity and the 
order parameter that breaks time-reversal symmetry}
\author {Valentin Stanev}
\affiliation {Institute for Quantum Matter and Department of Physics \& Astronomy, The Johns Hopkins University, Baltimore, MD 21218}
\author {Zlatko Te\v sanovi\' c}
\affiliation {Institute for Quantum Matter and Department of Physics \& Astronomy, The Johns Hopkins University, Baltimore, MD 21218}

\date {\today}

\begin{abstract} 
We consider a model of multiband superconductivity, inspired 
by iron pnictides, in which three bands 
are connected via repulsive pair-scattering terms. 
Generically, three distinct superconducting states arise within such a model. 
Two of them are straightforward generalizations 
of the two-gap order parameter while
the third one corresponds to a time-reversal symmetry breaking order
parameter, altogether absent within the two-band model. Potential observation
of such a genuinely frustrated state would be a particularly vivid manifestation
of the repulsive interactions being at the root of 
iron-based high temperature superconductivity.
We construct the phase diagram of this
model and discuss its relevance to the iron pnictides 
family of high temperature superconductors. 
We also study the case of the Josephson coupling between
a two-band $s'$ or $s\pm$ superconductor and a 
single-gap $s$-wave superconductor, and the associated phase diagram.   

\end{abstract}
\maketitle


\section{Introduction}
  The discovery of a new high-temperature superconducting family of iron-based materials
\cite {LaOFeAs, SmOFeAs, LaOFeAshd,CeOFeAs, PrOFeAs,BaFe2As2,SmFe2As2, 
SmFe2As2coex,BaFe2As2coex} and the subsequent developments 
have brought the question of multi-band superconductivity 
to the forefront of the condensed matter research. First 
discussed fifty years ago, this problem had remained somewhat 
obscure until iron pnictides, with their multi-band Fermi surfaces, made its understanding
an intellectual imperative. 
Following the initial discovery, an order parameter 
based on a two-band model was proposed as a likely possibility \cite{Mazin}. 
Soon thereafter,  this so-called extended $s$-wave (or $s\pm$ or $s'$) superconducting
state has been shown to be favored by the multiband electron dynamics
of iron pnictides, both within a random phase approximation 
(RPA) type picture \cite{Kuroki, EPL, Graser}
and in various renormalization group based approaches \cite{Chubukov, Vlad, Fa Wang}
-- as well as arising from a strongly correlated local limit \cite{Bernevig, Qimiao} --
and is currently viewed as the most plausible superconducting state for these compounds. 

  The first theoretical studies of a multi-band superconductivity \cite{Suhl, Moskalenko} 
were a straightforward generalization of the BCS 
theory, with gap equations for several bands and attractive interactions. 
The most interesting result was that the (two) 
superconducting gaps $\Delta^i(T)$ do not necessarily follow the 
single-gap BCS temperature dependence. Soon, however,
 it was realized \cite{Kondo} that the two-band model brings something 
conceptually new - superconductivity can be enhanced even by purely
repulsive interband interaction. This requires a relative minus 
sign between the gaps on different portions of the multiply connected
Fermi surface, while otherwise retaining an overall $s$-wave symmetry.
In this way, it was argued in Ref. \onlinecite{Kondo}, the electron-phonon
superconductivity in transition metals could receive an additional
boost from the Coulomb repulsion driven resonant pair scattering
between the broad s or p bands and narrow d bands at the Fermi level.

The above conceptual novelty, however, extends much deeper
than anticipated in Ref. \onlinecite{Kondo}. The purely electronic interactions
could, in principle, produce superconductivity even in the {\em absence}
of any phonon-mediated attraction. The superconductivity in this
case would arise solely through the resonant pair 
scattering between the two bands and both, or more as the case may be, of these
bands could be narrow d or even f bands. 
This promising mechanism for achieving 
high-temperature superconductivity -- using 
purely electron-electron interactions with cut-off 
of order of Fermi energy instead of Debye frequency, and 
thus potentially much higher transition temperature $T_c$ --  remained, however, largely 
ignored for the next fifty years. The reason is, 
basically, that the conditions in real materials are less than
favorable. For $s'$-state to be operational the 
superconductivity-driving interband pair scattering has to be 
stronger than the superconductivity-suppressing intraband repulsion 
(most commonly they both come from the screened 
Coulomb interaction in metals). This is unlikely for at least 
one reason: the interband interaction usually involves higher 
momentum transfer, bands typically being well-separated in the $k$-space, and is 
generically smaller. Thus, the sign-changing order parameter was considered unrealistic.

This perception changed last year, with 
the advent of iron pnictides. At least for 
those members of this family that exhibit
 the highest $T_c$s, a  nodeless multi-gap order parameter -- with some ARPES experiments seeing as many as four different gaps \cite{Ding} -- appears firmly established. The conventional electron-phonon interaction seems too weak to explain $T_c$ as high as $57~K$ (although some highly unconventional strong electron-phonon coupling 
still remains a remote possibility). This state of affairs makes the purely repulsive
electronic interaction as the source of superconductivity and the $s'$-wave state in particular very appealing, even though it is still not entirely clear how the 
generic repulsion problem, described in the previous paragraph, can be overcome. 
There are some very promising studies in this direction, based on the renormalization
group arguments and the peculiar band 
structure of these compounds \cite{Chubukov, Vlad, Fa Wang}, 
suggesting a plausible route to this superconducting state. 

   Many theoretical studies so far have used some variation of the 
two-band model. A number of useful results were derived 
and valuable insight was gained within this simplified picture
\cite{twobands}. The real materials, 
however, are more complex, and some tight-binding representations
of iron-pnictides  \cite{Vlad} indicate that typically 
three bands -- one electron and two hole-like -- are those most
strongly coupled in the pair-scattering channel (see also Ref.
\onlinecite{Chubukovii}). All this adds some urgency to the study of multiband 
superconductivity with three or four bands. In this paper, 
we concentrate on a rather generic three-band model with repulsive interactions. 
The main question we are interested in
is "Is something conceptually new emerging 
from this increase of the number of the bands?". 
The answer is "Yes,'' despite the fact that the gap equations themselves have the appearance 
of straightforward generalization of the two-band case. The reason for this
is the frustration which the additional band introduces into the
problem. 

To develop some intuitive understanding of the model let us 
start with an effective two-band situation. We ignore the intra-band interaction and consider only identical bands. If the coupling with the third band is 
negligible, there are two gaps $\Delta^1= - \Delta^2$ and the overall
magnitude is determined by the standard BCS relation. If we now introduce coupling to the additional band there are several possibilities. The system can stay in a two-gap state
-- now there are three such states -- and keep the remaining
band (nearly) gapless. In that sense, the interactions between the bands are frustrated,
i.e. with such superconducting order one of the bands will not achieve what would otherwise
be its natural gapped state. As suggested previously \cite{Ng}, there is 
also a possibility for a new superconducting order parameter which compromises 
between the different frustrated two-gap order parameters. We show that 
this indeed happens within our microscopic model and intrinsically 
{\em complex} superconducting order parameter emerges naturally (of course,
there is always an arbitrary overall phase). Such superconducting
state spontaneously breaks the time-reversal symmetry and minimizes 
the ground state energy for a range of coupling constants, which we determine
below. For reader's benefit, we note here that an interesting and different possible 
time-reversal symmetry breaking order parameter, involving $s$- and $d$-wave coexistence,
was considered in the context of pnictides in Ref. \onlinecite{SCZhang}. Finally, if one of the Josephson-like
couplings between the bands is much smaller than the other two, one intuitively
expects that yet another form of the order parameter will appear: three 
gapped bands with a relative minus sign between the stronger-coupled ones. 
We show below that all of these possibilities are realized in different parts of 
the phase diagram of the microscopic model.     

\section{The model and its gap equations}

We start with a Hamiltonian which is a straightforward 
generalization of the single-band BCS theory. Our model therefore bears 
all the birth-marks of the original -- restriction to weak coupling, 
omission of many details concerning band structure and dynamics of 
interactions, etc. -- but shares some of its virtues as well: 
broad generality and simplicity which allows for analytic treatment. 
More realistic considerations would basically 
result in various quantitatively important but conceptually
straigthforward ``decorations'' of this simplified Hamiltonian, 
which we now write down in its reduced form:
\be    
\mathscr{H} - &\mu N_{op}& = \sum_{i,\bf{k},\sigma} \xi^{(i)}_{\bf{k}} c^{(i)\dagger}_{\bf{k}\sigma} c^{(i)}_{\bf{k}\sigma}  + \nonumber \\ 
&+& \sum_{i, j,\bf{k, k'}} G^{(ij)}_2 c^{(i)\dagger}_{\bf{k}\uparrow} c^{(i)\dagger}_{\bf{-k}\downarrow} c^{(j)}_{\bf{k'}\uparrow} c^{(j)}_{\bf{-k'}\downarrow} 
+ h.c. ~,
\label{Hamiltonian}
\ee
where the $i$ and $j$ are band indexes (they run from $1$ to $3$) and for the moment 
we assume $G^{(ii)}_2=0$ (i.e. {\em no intra-band} pair-scattering). 
This simplifies the calculations significantly and is justified by the following reasoning. If we are to include the intra-band terms, there will be a finite critical strength for $G^{(ij)}_2(k, k')$, below which superconductivity cannot exist (for repulsive interactions). Above this threshold, when the superconducting state is {\em already present}, the intra-band terms are irrelevant for the structure of the order parameter, which is entirely determined by the inter-band pair-scattering. This argument, however, has to be applied carefully (see below).
The Josephson-like term $G^{(ij)}_2(k, k')$ is separable and has the usual square-well form. We also assume identical parabolic 
two-dimensional ($2D$) bands. As we will see, in the gap equations it does not really matter whether we use hole or electron bands or some combination. So our results apply for 
all of these cases, although, of course, the precise dynamics that produces
superconductivity in iron pnictides is most likely directly tied to its semimetallic
character and the presence of both hole and electron bands at the Fermi level.
This general nature of our results is the consequence of the simplified
model and the relatively restricted set of question we are asking (for example,
our focus is on the structure of the order parameter). 
We can think of Eq. (\ref{Hamiltonian}) as describing effective 
low-energy theory and $G_2^{(ij)}$ as phenomenological parameters 
in which we have stored all the details about the realistic 
band structure and the high-energy processes. 

  Now, after we define mean field averages
\be
\Delta_{\bf k}^{i} = - \sum_{j\neq i, \bf{k'}} G_2^{(ij)}(k,k')\langle c^{j}_{-\bf{k'} \downarrow}  c^{j}_{\bf{k'} \uparrow}\rangle  
\ee 
and introduce Bogoliubov-transformed fermionic operators, by 
using the properties of $G^{(ij)}_2$ and following the usual algebra, 
we get a set of three gap equations \cite{Suhl, Moskalenko}  
\be 
\Delta^{i} = - \sum_{j \neq i }G_2^{(ij)} N^{j} T \sum_{\omega_n} \int_0^{\omega_C} d\xi \frac{\Delta^{j}}{(\omega_n)^2 +(E^{j})^2}~,
\label{gapeq}
\ee  
where $E^{i} = \sqrt{(\xi^{i})^2 + (\Delta^{i})^2}$, $\omega_n$ 
are the fermionic Matsubara frequencies and $\omega_C$ is high-energy cut-off. 
 
  Despite the apparent similarity with the single-band BCS theory, these non-linear
gap equations are considerably more involved, and do not allow 
for analytic solutions in the general case, even at $T=0$. To achieve some progress,
we need to simplify the model even further. Let us start with two bands, coupled via $G_2^{(23)}$, and then gradually turn on their couplings with a third band. 
We also assume these two new coupling constants $G_2^{(12)}$ 
and $G_2^{(13)}$ to be equal. Thus, we reduce the three generally 
different couplings to two and introduce dimensionless constants:
\be 
\lambda^{(12)} &=& N(0) G_2^{(12)},\  \lambda^{13} = N(0) G_2^{(13)},\  \lambda^{(23)} = N(0) G_2^{(23)}; \nonumber\\
\lambda^{(12)}&=&\lambda^{(13)} \equiv \lambda,\ \  \lambda^{(23)}\equiv \eta;\ \ \lambda, \eta > 0 ~,\nonumber
\ee      
where we have denoted the density of states (DOS) on the Fermi level as $N(0)$ (identical bands!).
With these simplifications, we are finally ready to make some analytic progress
and gain some insight of the physics of our model.

\section{Critical temperature and $T\approx T_c$ region} 

  We now proceed by linearizing Eqs. (\ref{gapeq}) in the region $T\approx T_c$ and $|\Delta|\ll T$. The problem then reduces to finding the eigenvectors and eigenvalues of a $3\times 3$ matrix. The possible order parameters are proportional to the eigenvectors, and the eigenvalues determine $T_c$. In this case the Eqs. (\ref{gapeq}) are equivalent to
\be
  I \left(
\begin{matrix}
  0     &\lambda  &\lambda  \\ 
\lambda &     0   &\eta    \\ 
\lambda & \eta    &0 
\end{matrix}
\right)
\left(
\begin{matrix}
  \Delta^1 \\ 
  \Delta^2\\
  \Delta^3 
\end{matrix}
\right)= -
\left(
\begin{matrix}
  \Delta^1 \\ 
  \Delta^2\\
  \Delta^3 
\end{matrix}
\right),
\ee
with $I = \gamma\ln{(2\omega_C/\pi T_c)}>0$ ($\gamma$ is the Euler constant). Solving 
this matrix equation gives us three real eigenvalues
\be
\delta_{i} = -I \eta,\ \frac{I}{2}(\eta - \sqrt{8\lambda^2 + \eta^2}),\  \frac{I}{2}(\eta + \sqrt{8\lambda^2 + \eta^2}),  
\ee
and their corresponding eigenvectors 
\be
\Delta_{i} \propto
\left(
\begin{matrix}
   0\\ 
  -1\\
   1 
\end{matrix}
\right),\ 
\left(
\begin{matrix}
  - \frac{\eta + \sqrt{8\lambda^2 + \eta^2}}{2 \lambda} \\ 
   1 \\
   1 
\end{matrix}
\right),
\left(
\begin{matrix}
  - \frac{\eta - \sqrt{8\lambda^2 + \eta^2}}{2 \lambda}\\ 
   1\\
   1 
\end{matrix}
\right).
\ee

For $\lambda, \eta>0$ there are two negative eigenvalues and accordingly two possible order parameters.  If we fix $\eta$ and gradually increase $\lambda$ from zero it is easy to see that eigenvalues $\delta_1$ and $\delta_2$ cross at the point $\eta=\lambda$. The $\tilde{\Delta}_1$ is obviously the two-gap solution and has higher $T_c$ for $\lambda<\eta$. The other possibility is a three-gap superconductor with a relative minus sign between those
bands that experience stronger coupling. Thus, the order parameter can be chosen 
to be real along the {\em entire} $T_c$ line. Precisely at the crossing point, 
the eigenvalues are degenerate and there any superposition of the two eigenvectors is also a legitimate order parameter. This degeneracy is a consequence of linearizing Eqs. (\ref{gapeq}) 
and it leads to the possibility of complex $\tilde{\Delta}$, with non-trivial 
phase difference between the components. One example is the $\tilde{\Delta} \propto \{ 1, e^{\frac{2 i\pi}{3}},e^{-\frac{2 i\pi}{3}} \}$ -- the Ginzburg-Landau theory of this particular state was constructed and studied in Ref. \onlinecite {Agterberg}. 

  Once $T$ is below $T_c$ and one enters
the superconducting state, we expect the complex order parameter to emerge as a competitor to the real one within a {\em finite} region, as opposed to a point at $T=T_c$. Because of the $2\leftrightarrow 3$ symmetry in the gap equations, we will look for solutions that satisfy the condition $|\Delta^2| = |\Delta^3|$. We can write 
the ordinary two- and three-gap order parameter as
\be 
\tilde{\Delta}_1 = \{\  0,\  - 1\ , 1 \} \Xi,\ \  \tilde{\Delta}_2 = \{\  - \theta\ , 1,\  1 \}\Lambda
\label{ROP}
\ee 
and introduce intrinsically complex, time-reversal symmetry 
breaking order parameter of the form 
\be
\tilde{\Delta}_3 = \{ -\kappa,\  e^{i \varphi},\   e^{- i \varphi} \}\Omega.
\label{FOP}
\ee  
In Eqs. (\ref{ROP}) and Eq. (\ref{FOP}) $\Xi$, $\theta$, $\Lambda$, $\kappa$, $\Omega$ and $\varphi$ are all real variables that parametrize the order parameters on the different bands and are to be determined self-consistently by solving the gap equations.

\begin{figure}[htb]
\includegraphics[width=0.35\textwidth]{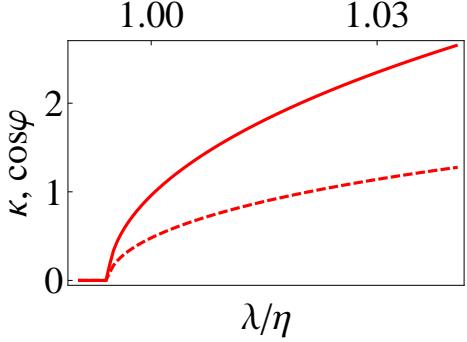}
\caption{ (Color online) Plot of $\kappa$ (solid line) and $\cos\varphi$ (dashed line) for $\tilde{\Delta}_3$ at $T= 0.95 T_c$. This OP exists as a distinct solution only at a vicinity of the degenerate point $\lambda=\eta$. On the left it merges with $\tilde{\Delta}_1$ and on the right crosses $\tilde{\Delta}_2$ at $cos\varphi = 1$ and disappears. The interval is asymmetric with respect to the degenerate point.} 
\label {KappaPsi}
\end{figure}

  To proceed analytically we expand the Eqs. (\ref{gapeq})
for $(T_c -T)/T_c$ to second order in the magnitude of the order parameter. The two-gap solution obviously follows the BCS behavior:
\be
\frac{1}{\eta} = \gamma \ln(\frac{2 \omega_c}{\pi T}) - \beta_0\frac{\Xi^2}{T_c^2}.
\label{geqT1} 
\ee 
where we have simplified the notation by introducing a new constant $\beta_0 = 7\zeta(3)/8\pi^2 $.
 The three-gap order parameters have more complicated behavior:
\be
&&\lambda \theta(1 - \theta^2)\ln(\frac{2 \omega_c}{\pi T})+ \frac{\theta^4}{2} - \frac{\eta \theta}{2 \lambda} - 1 =0,\nonumber \\   &&\frac{\theta}{2 \lambda} = \gamma \ln(\frac{2 \omega_c}{\pi T}) - \beta_0 \frac{\Lambda^2}{T_c^2},
\label{geqT2} 
\ee
and
\be
&&\lambda \kappa(1 - \kappa^2)\ln(\frac{2 \omega_c}{\pi T}) + \frac{\lambda \kappa^3}{ \eta} - \frac{\eta \kappa}{ \lambda} =0,\nonumber \\&& cos\varphi = \frac{\eta \kappa}{2 \lambda} ;\  \ \ \frac{1}{\eta} = \gamma \ln(\frac{2 \omega_c}{\pi T}) - \beta_0\frac{\Omega^2}{T_c^2}.
\label{geqT3} 
\ee
  We now solve these equations numerically and obtain $\tilde{\Delta}_1$, $ \tilde{\Delta}_2$ and $\tilde{\Delta}_3$.
 The time-reversal symmetry breaking (TRSB) solution exist only in the
narrow interval $\lambda \in (\lambda_{c1}, \lambda_{c2})$, where $\lambda_{c1}, \lambda_{c2} \rightarrow \eta$ for $T\rightarrow T_c$. For $\lambda$  smaller than $\lambda_{c1}$, the complex order parameter reduces to the two-gap one, while
for the coupling parameter bigger than $\lambda_{c2}$ only the trivial solution exist for $\tilde{\Delta}_3$ (see Fig. (\ref{KappaPsi})).
 
\begin{figure}[htb]
\includegraphics[width=0.35\textwidth]{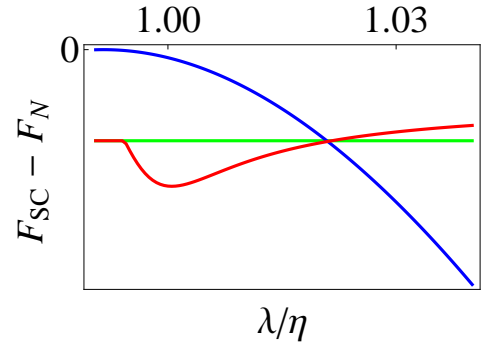}
\caption{ (Color online) Comparison between the $\mathcal{F}_{SC} - \mathcal{F}_{N}$  for $\tilde{\Delta}_1$ (green), $\tilde{\Delta}_2$ (blue)  and $\tilde{\Delta}_3$ (red). We show calculation on the right side of the $\lambda= \eta$ point for $T= 0.95 T_c$. The complex OP minimizes the free energy in a small interval. The transition between $\tilde{\Delta}_3$  and $\tilde{\Delta}_2$ is discontinuous.} 
\label {FreeEnergydif}
\end{figure}

  Using Eqs. (\ref{geqT1}), (\ref {geqT2}) and (\ref {geqT3}) we can proceed to calculate all the thermodynamic quantities of interest. We follow a simple route - solve Eqs. (\ref{geqT2}) for $\theta(T, T_c)$ and Eqs. (\ref{geqT3}) for $\kappa(T, T_c)$,  $\varphi(T, T_c)$ (their structure allows it). After that we construct three single-variable Ginzburg-Landau theories (assuming uniform solutions), minimize the free energies and compare the results:
\be
 \mathcal{F} = \alpha_{i} |\psi_i|^2 + \frac{\beta_{i}}{2} |\psi_i|^4, \ \  \mathcal{F}_{min} = - \frac{\alpha_{i}^2}{2\beta_{i}},
\label{GL}
\ee    
 where $|\psi_1| = \Xi, |\psi_2|= \Lambda$ and $|\psi_3|= \Omega$. To do that we write the interaction part of the Hamiltonian as 
\be
\mathscr{H}_{int} = \sum_{i, j,\bf{k, k'}} G^{(ij)}_2 c^{(i)\dagger}_{\bf{k}\uparrow} c^{(i)\dagger}_{\bf{-k}\downarrow} d^{j} 
+ h.c. 
\label{intH}
\ee
and in Eq. (\ref {intH}) we have introduced auxiliary mean field averages $d^{i}  = - \sum_{\bf{k}} \langle c^{i}_{-\bf{k} \downarrow}  c^{i}_{\bf{k} \uparrow}\rangle  $. 
Using the definitions of $\{\Delta^{i}\}$ we can write equations for $d^{i}$:
\be
\Delta^1 &=& \frac{\lambda}{N(0)}(d^1 + d^2),\ \Delta^2 = \frac{1}{N(0)}( \lambda d^1 + \eta d^3),\nonumber\\ \Delta^3 &=& \frac{1}{N(0)}(\lambda d^1 + \eta d^2),  \nonumber 
\ee 
which can be solved for $\{d^{i}\}$:
\be
d^1 &=& N(0)\frac{ -\eta \Delta^1 + \lambda(\Delta^2 + \Delta^3)}{2 \lambda^2}, \nonumber\\d^2 &=& N(0) \frac{ \eta \Delta^1 + \lambda(- \Delta^2 + \Delta^3)}{2 \lambda \eta}, \nonumber\\
d^3 &=& N(0) \frac{ -\eta \Delta^1 + \lambda( \Delta^2 - \Delta^3)}{2 \lambda \eta}.  \nonumber 
\ee 
Now we can obtain connected-diagrams expansion for the free energy in orders of $d^i$.  The second and the forth-order terms come from the expressions
\be
\mathcal{F}_{2} \propto \int^{1/T}_0d\tau_1 \int^{1/T}_0d\tau_2 \langle T_{\tau} \mathscr{H}_{int}(\tau_1) \mathscr{H}_{int}(\tau_2)\rangle, \nonumber\\
\mathcal{F}_{4} \propto \int^{1/T}_0d\tau_1..\int^{1/T}_0d\tau_4 \langle T_{\tau} \mathscr{H}_{int}(\tau_1) ..\mathscr{H}_{int}(\tau_4)\rangle. \nonumber
\ee 
Using Eq. (\ref{intH}) for $\mathscr{H}_{int}$, properties of the electron Greens' functions $\mathscr{G}^{ij} \sim\delta_{ij}$ and expressing the $d^i$'s via the $\Delta^i$'s we eventually get $\mathcal{F}$ in the form of Eq. (\ref{GL}).
We have to calculate the pre-factors so we can compare the different order parameters. Expressions for $\alpha_1(T, T_c)$, $\alpha_2(T, T_c, \theta))$, $\alpha_3(T, T_c, \kappa, \varphi)$,  $\beta_1(T, T_c)$, $\beta_2(T, T_c, \theta))$ and $\beta_3(T, T_c, \kappa, \varphi)$ are straightforward but tedious to obtain and very unwieldy, so we just report the results for $ \mathcal{F}$. Far from the symmetric point $\lambda= \eta$ the solution with highest $T_c$ remains stable. On Fig. (\ref{FreeEnergydif}) we show the comparison between different $ \mathcal{F}$'s for $T= 0.95 T_c$  at the vicinity of the symmetric point  - the different solutions which are degenerate at $T_c$ split, and the complex order parameter has lowest free energy. This, however, remains true only in a relatively small interval around the line $\lambda=\eta$, which as $T\rightarrow T_c$ reduces to a point. This interval is asymmetric and 
considerably smaller on the $\lambda<\eta$ side. The different phases appear to be divided by a continuous phase transition on the left and a 
first order phase transition on the right side.      
     
 
\section{Low-temperature region} 
  Now we will concentrate on $T=0$ properties of the model. To distinguish the parameters from the finite-temperature case we use subscript "$0$". To find $\Xi_0$, $\theta_0$, $\Lambda_0$, $\kappa_0$ and $\Omega_0$  we have to solve $T=0$ version of Eq. (\ref{gapeq}).  

 The two-gap solution leads to identical gap magnitudes which obey the BCS relation:
\be
\Xi_0 \approx 2 \omega_C e^{- 1/\eta}.
\ee  
Since $\lambda$ does not enter the gap equation for $\tilde{\Delta}_1$, it is always a solution, irrespective of the ration $\eta/\lambda$.

 For the real three-gap solution $\tilde{\Delta}_2$ there are two unknowns to determine. We get the following equations for them:
\be
&\theta_0^2& - \frac{\eta}{\lambda} \theta_0 - 2\theta_0 \ln{\theta_0} - 2 = 0, \nonumber \\
&\Lambda_0& = 2 \omega_C e^{- \theta_0/2\lambda}.
\label {R3gap}
\ee
Eqs. (\ref{R3gap}) can be solved numerically and there is always a non-zero solution for $\theta_0$. That means that $\tilde{\Delta}_2$ is solution for all (non-zero) values of $\lambda$. Looking at $\Lambda_0$, however, we see that it is strongly suppressed for $\lambda \rightarrow 0$, which is to be expected since at $\lambda = 0$ the only non-trivial solution is $\tilde{\Delta}_1$.
 
\begin{figure}[htb]
\includegraphics[width=0.35\textwidth]{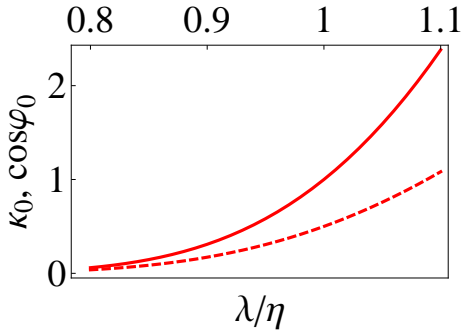}
\caption{ (Color online) Plot of $\kappa_0$ (solid line) and $\cos\varphi_0$ (dashed line) for $\tilde{\Delta}_3$ at $T=0$. This OP exists only for $\lambda\in(0,\lambda_{c0})$ where $\lambda_{c0}>\eta$.} 
\label {KappaPsi0}
\end{figure}

  For the complex three-gap order parameter we get equations
\be
\lambda \kappa_0 \ln(\kappa_0) - \kappa_0\left(\frac{\lambda^2 - \eta^2}{\lambda\eta} \right)= 0,\nonumber\\
\cos\varphi_0 = \frac{\eta \kappa_0}{2 \lambda};\ \  \Omega_0 = 2 \omega_C e^{- 1/\eta}.
\label {F3gap}
\ee
 These equations have non-trivial solutions for $\lambda\in(0,\lambda_{c0})$ where $\lambda_{c0}>\eta$ (see Fig. (\ref{KappaPsi0})). At the point  $\lambda = \eta$ the order parameter has the completely symmetric form  
\be
\tilde{\Delta}^{symm}_3 = \{ 1,\   e^{\frac{2i\pi}{3}},\   e^{-\frac{2i\pi}{3}} \} \Omega_0,
\ee  
which is easy to understand if we consider the gap equations. At this point $\Xi_0 =\Omega_0$.

  Now we want to see which order parameter is the actual ground state for different $\lambda$. We calculate the difference between the superconducting and the normal state energies for the different $\tilde{\Delta}$:
\be
 \mathcal{E}_{SC} - \mathcal{E}_{N} = \langle \Psi_{\tilde{\Delta}}|\mathscr{H} - \mu N_p | \Psi_{\tilde{\Delta}}\rangle  - \langle \Psi_{FS}|\mathscr{H} - \mu N_p|\Psi_{FS} \rangle. \nonumber
\ee
The normal and superconducting state kinetic energies are respectively:
\be
\mathcal{KE}_{N}& =&  \sum_{i, k<k_F} 2\xi^{i}_{\bf{k}}, \\ \nonumber
\mathcal{KE}_{\tilde{\Delta}}& = & \sum_{i, \bf{k}} \left( \xi^{i}_{\bf{k}} - \frac{(\xi^{i}_{\bf{k}})^2}{E^{i}_i}\right)\nonumber.
\ee  
  Converting the sum into an integral gives           
\be 
\mathcal{KE}_{\tilde{\Delta}}-\mathcal{KE}_{N} = 2 \sum_{i} N(0)\int_0^{\omega_C} \left( \xi^{i}_{\bf{k}} - \frac{(\xi^{i}_{\bf{k}})^2}{E^{i}}\right) \approx \nonumber\\  N(0)\sum_{i}\left( |\tilde{\Delta}^{i}|^2 \sinh^{-1}\left(\frac{\omega_C}{|\tilde{\Delta}^{i}|}\right) -\frac{1}{2}|\tilde{\Delta}^{i}|^2\right). 
\label {KE}
\ee   
The mean field average of the potential energy in the normal state is zero, and for calculation in the superconducting state we again use $d^{i}$. Then the potential energy can be written as
\be
\mathcal{PE}_{\tilde{\Delta}_i} - \mathcal{PE}_{N}  = \lambda d^{1*} d^2 + \lambda d^{1*} d^3 + \eta d^{2*} d^3 + h.c
\label {PE}.
\ee

 Using the expressions for $d^{i}$ in the potential energy formula we get
\be
\mathcal{PE}_{\tilde{\Delta}} &-& \mathcal{PE}_{N}  = N(0) \frac{ - \eta |\Delta^1|^2 + \lambda^2(- |\Delta^2|^2 + |\Delta^3|^2) }{\lambda^2 \eta} \nonumber\\ &+& N(0)\frac{ \lambda \eta (\Delta^1(\Delta^{2*} + \Delta^{3*}) + \Delta^{1*}(\Delta^{2} + \Delta^{3}) )}{\lambda^2 \eta}. \nonumber
\ee   

  Combining Eqs. (\ref{KE}) and (\ref{PE}) we can compute the energies for the different possible ground states and compare them. Let us start with $\tilde{\Delta}_1$ - since $\Xi_0$ follows the BCS behavior we get the standard result, multiplied by $2$ (two bands):
\be
\mathcal{E}_{\tilde{\Delta}_1} - \mathcal{E}_{N}  = - N(0) \Xi_0^2. 
\ee    

  Similar calculations for $\tilde{\Delta}_2$ and $\tilde{\Delta}_3$ give the energy difference as a function of $\theta_0$ and $\Lambda_0$, or $\kappa_0$ and $\Omega_0$:
\be
\mathcal{E}_{\tilde{\Delta}_2} - \mathcal{E}_{N} = -  N(0) \left( 1 +  \frac{\theta_0^2}{2} \right)\Lambda_0^2,
\ee
\be
\mathcal{E}_{\tilde{\Delta}_3} & - & \mathcal{E}_{N} = \\
& & - N(0) \left( 1 - \frac{2}{\eta} + \frac{\kappa_0}{2}(1 - \frac{2}{\eta} + 2 \ln\kappa_0 )\right)\Omega_0^2 -\nonumber\\
& & N(0) \left( \frac{\eta \kappa_0^2 + 4\lambda \kappa_0 \cos\varphi_0}{2 \lambda^2} + \frac{ 2 \sin^2\varphi_0}{\eta} \right)\Omega_0^2. \nonumber
\ee

\begin{figure}[htb]
\includegraphics[width=0.35\textwidth]{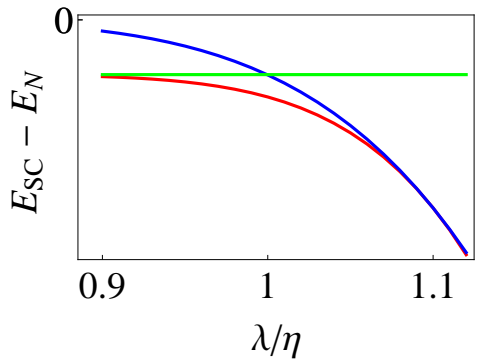}
\caption{ (Color online) Comparison between the $\mathcal{E}_{SC} - \mathcal{E}_{N}$  for $\tilde{\Delta}_1$ (green), $\tilde{\Delta}_2$ (blue)  and $\tilde{\Delta}_3$ (red). The first one is never a ground state for $\lambda\neq 0$. The energies for $\tilde{\Delta}_2$ and $\tilde{\Delta}_3$  merge for some $\lambda_{cr} > \eta$.} 
\label {Energydif}
\end{figure}

   
  Now we are able to compare the different solutions - the result is shown on Fig. (\ref{Energydif}). On the left side of $\eta$ we see that the time-reversal symmetry breaking order parameter is the ground state. It converges from below to the two-gap solution as $\lambda, \theta_0 \rightarrow 0$. At the symmetric point $\lambda=\eta$ we can use the BCS result for both solutions and $ \mathcal{E}_{\tilde{\Delta}_3} - \mathcal{E}_{N} = 3/2(\mathcal{E}_{\tilde{\Delta}_1} - \mathcal{E}_{N})$ (three vs two gaps). On the right side there is a Quantum Phase Transition at some $\lambda_{cr}> \eta$, where the complex and the real three-gap states merge ($\cos\varphi=1$). Beyond this point $\tilde{\Delta}_3$ ceases to exist.

\section{Phase diagram} 
\begin{figure}[htb]
\includegraphics[width=0.4\textwidth]{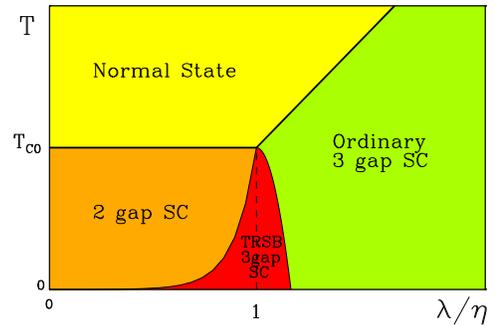}
\caption{ (Color online) Suggested phase diagram of the three-band model. There are three possible superconducting (SC) order parameters (OPs). The line separating the TRSB and real three-gap OP is most likely first order phase transition line.} 
\label {phasediagram}
\end{figure}

  On the basis of the above results we suggest that our model has
 the phase diagram depicted in Fig. (\ref{phasediagram}). There are three 
superconducting order parameters, stable in different regions, separated by two critical lines. On the left, different superconducting states appear to be separated by continuous transition and on the right by first order one at finite temperature, and continuous one at $T=0$. 
There is a possibility of observing two different superconducting states in a single system and the transition between them, tuned by the temperature. In the context of this phase diagram we consider the case of iron pnictides - bands $2$ and $3$ can be thought as the hole bands at the $\Gamma$ point, which are strongly coupled to one of the electron bands at the  $M = (\pi, \pi)$ point \cite{Vlad}. If the renormalization group arguments apply, $G_2^{e_1h_i}$ are enhanced and $G_2^{h_1h_2}$ is suppressed by the same high-energy electron-hole processes. Then the appropriate regime for the pnictides is $\lambda > \eta$, on the right side of our diagram. The existence of time-reversal symmetry breaking order parameter is not excluded, but is unlikely for the optimally doped compounds, given the relative narrowness in which it is stable for $\lambda > \eta$. However it may be present in the overdoped materials, for which 
the inter-band interactions are suppressed, due to the 
significant deviations from perfect nesting. This complex order parameter, in general, entails the existence of local magnetic fields at edges  and around impurities, and likely domain structure \cite{Sigrist}. These effects may provide the best way of observing such state and its broken symmetry.         

Before we proceed, let us comment on two obvious deficiencies in our model, which seemingly prevent us from applying the results we have derived thus far to the iron pnictides. First, we have completely neglected the intra-band pairing terms. Once these terms are included the calculations become considerably more involved and it
is difficult to proceed short of pure numerics. However, we believe that our phase diagram is qualitatively correct even in
that case, since, as already explained, these terms do not play a role in determining the structure of the order parameter (provided that the superconductivity is still possible) and only change the numerical values of various results ($T_c$, for example). The validity of this argument is limited, however, and the intra-band terms have an important role to play in the case of several competing channels ($s$ and $d-$wave, for example) which are affected differently by these terms. There are several studies for pnictides suggesting such competition \cite{Kuroki,Chubukov2, Thomale}. But as long as the most isotropic superconducivity remains the leading instability, it will be realized without any mixing from the sub-leading channels (for $s$ and $d$ mixing see Ref. \onlinecite{sd}) and our results apply. In case the system is driven to a nodal state by the intra-band repulsion, the frustration due to the inter-band terms can again lead to a development of complex order parameter, but we leave this question for further studies. Second, we have restricted ourselves to a three-band model, whereas in pnictides there are generally four active bands participating in the superconductivity (see, for example, Ref. \onlinecite {Ding}). It is a valid question if adding another band will  
 completely suppress the complex order parameter. To address it let us remind the reader the tight-binding calculation\cite{Vlad}, which indicates that, for the case of pnictides, the pair-scattering terms between the second electron band (whose existence we have neglected) and the hole bands are at least an order of magnitude smaller that $G^{e_1h_i}$ (analogous to our $G^{13}_2$ and $G^{23}_2$ terms). Coupling between the electron bands, however, generically will be of the same order as $G^{h_1h_2}$ (our $G^{12}_2$).  This means, in practice, that the phase of the gap on the second electron band will (almost) entirely depend on the gap on the other \emph{electron} band. Thus the relative phase between the gaps opening on the hole and the electron bands will still be determined by the three-band calculation. These (somewhat naive) arguments allow us the hope that our model, despite its simplicity and numerous assumptions, is relevant for the iron pnictides.      
  
 One condition of particular relevance to the pnictides is the condition for existence of superconductivity itself - in the two-band model it is $G^{12}_2 > U$, ($G^{11}_2=G^{22}_2\equiv U$ is the intra-band pairing). For the real three-band order parameter in the limit $\lambda \gg \eta$ this condition becomes $G^{\lambda}_2 > U/\sqrt{2}$, i.e. it is somewhat relaxed. 
  
\section{Josephson-coupled two-gap s\' \  state and single-gap s state}
\begin{figure}[htb]
\includegraphics[width=0.23\textwidth]{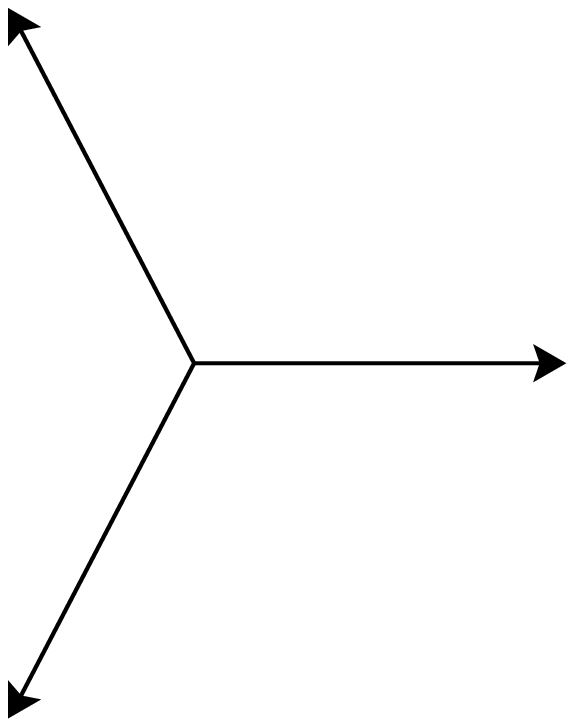}
\hfill
\includegraphics[width=0.23\textwidth]{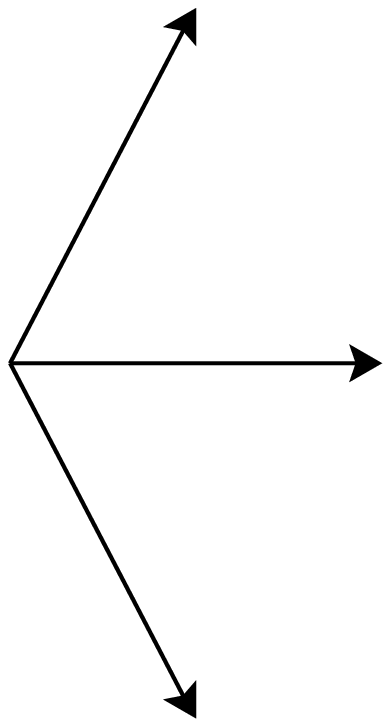}
\caption{ Schematic representation of the TRSB order parameter in the case of three positive inter-band couplings (left) and two negative and a positive inter-band couplings (right). The frustration is resolved in a different but related way.} 
\label {arrows}
\end{figure}

\begin{figure}[htb]
\includegraphics[width=0.35\textwidth]{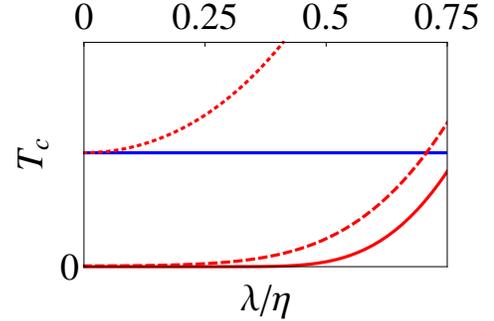}
\caption{ (Color online) Comparison of the $T_c$ for $\tilde{\Delta}_1$ (blue) and $\tilde{\Delta}_2$ (red) for $b=0$ (solid line), $b=0.5$ (dashed line) and $b=1$ (dotted line). $\eta$ is fixed. For small $\lambda$ the two-gap solution is the first to appear for $b<1$. At the crossing point there is degeneracy and complex $\tilde{\Delta}$ is possible.} 
\label {Tcrit}
\end{figure}

  We can use the model and the results derived so far to 
study a different problem - a two-gap $s'$ state coupled via 
Josephson junction to an ordinary $s$-wave superconductor. This is a situation
of real experimental relevance, in light of the recent experiments demonstrating
Josephson effect between Pb and an iron-pnictide superconductor
\cite{Greene}; a theoretical background is explored in
\cite{Linder, Ghaemi, Pedro}. 
The tunneling of Cooper pairs in this case 
would like to align the phases of the two (distinct) 
superconductors - we can model this by introducing negative coupling constant $\lambda \rightarrow - \lambda $ (and $|\lambda|\ll\eta $ - weak coupling). 
It does not take one much time to realize that the 
equations for this model can be made identical to 
the ones for the previous model by a single sign 
flip in the three-gap order parameters. For example,
the real solution $\tilde{\Delta}_2$ now becomes:
\be 
 \{\  - \theta\ , 1,\  1 \}\Lambda \rightarrow  \{\  \theta\ , 1,\  1 \}\Lambda.
\ee 
After this change the phase diagram is identical to that on Fig. (\ref{phasediagram}). The model is still frustrated, but the frustration is resolved in a different manner - the non-trivial phase angle now brings the $\eta$-coupled gaps closer (instead of further away) to the third gap. This is easy to see at the completely degenerate point $\lambda=\eta$ (Fig. (\ref{arrows})):  
\be
\tilde{\Delta}^{symm}_3 \rightarrow \{ 1,\   e^{\frac{i\pi}{3}},\   e^{-\frac{i\pi}{3}} \} \Omega.
\ee
 The model with negative Josephson junctions still does not give us two \emph {independent} superconductors. To achieve this we add intra-band attraction on the weakly-coupled band. Now even at $\lambda=0$ we have two different superconducting states - a single-gap $s$-wave and a $s'$-wave two-gap solutions (previously $\lambda$ solely was driving the superconductivity on the third band). The equation for $T_c$  becomes:
\be
  I \left(
\begin{matrix}
  -b \eta     &-\lambda  &-\lambda  \\ 
-\lambda &     0   &\eta    \\ 
-\lambda & \eta    &0 
\end{matrix}
\right)
\left(
\begin{matrix}
  \Delta^1 \\ 
  \Delta^2\\
  \Delta^3 
\end{matrix}
\right)= -
\left(
\begin{matrix}
  \Delta^1 \\ 
  \Delta^2\\
  \Delta^3 
\end{matrix}
\right),
\ee
where we have parametrized the intra-band attraction as a fraction $b$ of $\eta$. For the experimental set-up of a conventional low-temperature superconductor coupled to iron pnictide sample we expect $b < 1$. The eigenvalues and eigenvectors now are:
\be
\delta_i = -I \eta,\ \ \frac{I}{2}(\eta(1 - b) \mp \sqrt{8\lambda^2 + \eta^2(1 + b)^2}),  
\ee
\be
\tilde{\Delta_i} \propto
\left(
\begin{matrix}
   0\\ 
  -1\\
   1 
\end{matrix}
\right),\ 
\left(
\begin{matrix}
   \frac{\eta(1 + b) \pm \sqrt{8\lambda^2 + \eta^2 (1 + b)^2}}{2 \lambda} \\ 
   1 \\
   1 
\end{matrix}
\right).
\ee  
Again there are two possible order parameters. The $T_{ci}$ curves (Fig. (\ref{Tcrit})) still cross but their crossing point is no longer at $\lambda = \eta$. It moves to the left, which is easy to understand - the $T_c$ line for the three-gap OP goes to a finite limit rather than zero for $\lambda \rightarrow 0$ (single-gap SC, courtesy of the non-zero $b$). 
  
\begin{figure}[htb]
\includegraphics[width=0.35\textwidth]{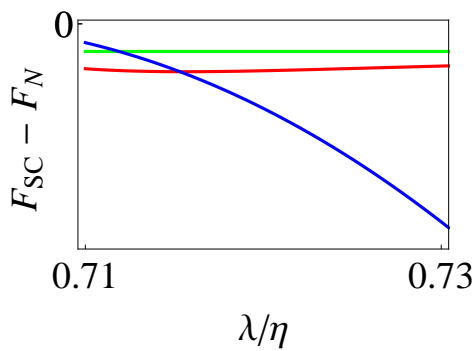}
\caption{ (Color online) Comparison between the $\mathcal{F}_{SC} - \mathcal{F}_{N}$  for $\tilde{\Delta}_1$ (green), $\tilde{\Delta}_2$ (blue)  and $\tilde{\Delta}_3$ (red) for $b=0.5$. We show calculation for $T= 0.95 T_c$ at the vicinity of the $T_{ci}$ crossing point $\lambda\approx 0.71 \eta$. The interval for which the complex OP minimizes the free energy is smaller but still exist.} 
\label {FreeEnergydif2}
\end{figure}

  We again expand the gap equations in the vicinity of $T_c$ up to second order in the magnitudes of the superconducting gaps. The two-gap solution is unchanged, and the real three-gap solution equations become:
\be 
&&\left((\lambda \theta - \eta) - (\lambda\theta^3 - \eta )\frac{2 \lambda + b \eta \theta}{2 \lambda + b \eta \theta^3}\right)\ln(\frac{2 \omega_c}{\pi T}) - \nonumber\\&& (\eta - \lambda \theta^3)\frac{\theta}{2 \lambda + b \eta \theta^3} -1 =0, \nonumber \\
 &&\theta = \gamma (2 \lambda + b \eta \theta) \ln(\frac{2 \omega_c}{\pi T}) - (2 \lambda + b \eta \theta^3)\beta_0 \frac{\Lambda^2}{T_c^2}.
\ee
The complex order parameter equations are:
\be
&&\lambda \kappa(1 - \kappa^2)\ln(\frac{2 \omega_c}{\pi T}) + \frac{\lambda \kappa^3}{ \eta} - \frac{\eta \lambda \kappa}{ (b \eta^2 + \lambda^2)} =0,\nonumber \\&& cos\varphi = \frac{\eta \lambda \kappa}{2 (b \eta^2 + \lambda^2)} ,\  \frac{1}{\eta} = \gamma \ln(\frac{2 \omega_c}{\pi T}) - \beta_0\frac{\Omega^2}{T_c^2}.
\ee
For $b\rightarrow 0$ these equations reduce correctly to the inter-band-couplings-only case. We derive single-variable Ginzburg-Landau free energy, and then minimize it with respect to $\Lambda$ and $\Omega$. The comparison between the different solutions for $b=0.5$ is shown on Fig. (\ref{FreeEnergydif2}). The result is very similar to the $b=0$ case but the region in which the complex order parameter dominates is smaller. The reason is that the real three-gap solution's free energy is pushed down by the intra-band term. However, for $b\rightarrow 1$ the region again expands as the crossing point is pushed closer to $\lambda = 0$. For $b>1$ the real three-gap solution minimizes the $\mathcal{F}$ for all $\lambda$, at least for $T\approx T_c$.

\begin{figure}[htb]
\includegraphics[width=0.35\textwidth]{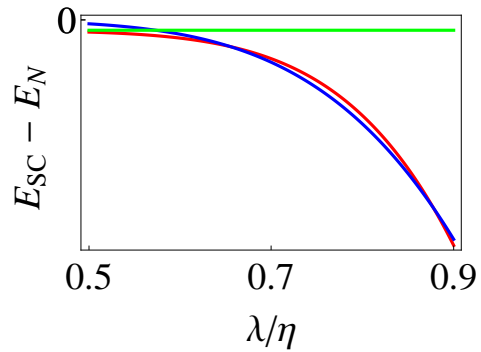}
\caption{ (Color online) Comparison between the $\mathcal{E}_{SC} - \mathcal{E}_{N}$  for the different order parameters. Here $b=0.5$. The energies for $\tilde{\Delta}_2$ and $\tilde{\Delta}_3$  cross for $\lambda_{cr}\approx 0.65$.} 
\label {Energydif2}
\end{figure}

  Now we discuss the $T=0$ line of the phase diagram. The gap equations for $\tilde{\Delta}_2$ and $\tilde{\Delta}_3$ are now:
\be
&\theta_0^2& - \frac{\eta}{\lambda} \theta_0(1 - b (1 + \eta \ln{\theta_0})) - 2\theta_0 \ln{\theta_0} - 2 = 0, \nonumber \\
&\Lambda_0& = 2 \omega_C e^{- \theta_0(1 + b \eta \ln{\theta_0})/(2\lambda + b \eta \theta_0)};\nonumber
\ee
and
\be
&&\lambda \kappa_0 \ln(\kappa_0) - \kappa_0\left(\frac{\lambda^2 - \eta^2}{\lambda\eta} \right) - b \frac{\eta}{\lambda} \kappa_0 (1 - \eta \ln{\kappa_0})= 0,\nonumber\\
&&\cos\varphi_0 = \frac{\eta \kappa_0}{2 \lambda}(1 - b + b\eta \ln{\kappa_0}),\ \  \Omega_0 = 2 \omega_C e^{- 1/\eta}.\nonumber
\ee
 We compare the energies for the different possible ground states on Fig. (\ref {Energydif2}). Consistent with the results from the $T_c$ region, the complex solution is still the ground state on the left of some $\lambda_{cr}$. However, because the energy of $ \tilde{\Delta}_2$ is pulled down for non-zero $b$, the transition a weakly first-order one and $\lambda_{cr}$ is on the left of the $T_{ci}$ crossing point.   
 
  With increase of $b$, $\lambda_{cr}$ moves to the left, but there is always a region (confined to lower and lower temperatures and smaller and smaller $\lambda$ as $b$ goes up) in which the complex solution is the preferred order parameter.    

\section{Conclusions}
  In summary, we have considered a simple microscopic model, with three bands coupled via repulsive pair-scattering interactions, which is relevant for the recently discovered 
iron-based family of high-temperature superconductors. We have 
constructed the phase diagram of this model and discussed its overall
features. Generally, we find
three possible superconducting order parameters, 
one of which breaks the time-reversal symmetry
in order to relax some of the frustration 
intrinsic to the three (or odd) band case. 
The conditions for such exotic state are rather strict and it
seems unlikely that this order parameter would be 
observed in the optimally doped iron pnictides. 
However, this state may be realistically present in overdoped samples,
if the doping is carefully tuned to the range of optimized frustration. 
While quantitative aspects of our results are bound to be sensitive to
the details of the band-structure and the accompanying
orbital character of each individual
iron-pnictide material -- the details which
are not part of our model -- the overall qualitative features reported in this
paper are expected to remain relatively universal.
Experimental observation of a time-reversal symmetry breaking 
superconducting state is perhaps the best we can hope for in linking 
an $s'$ superconductor to some broken symmetry and 
would represent arguably the strongest confirmation yet
of the basic picture which places the
repulsive, purely electron-electron interband
interactions at the heart of iron-based 
high-temperature superconductivity. 
Furthermore, we have also considered the case of Josephson-coupled 
two-band $s'$ SC and a single-gap $s$ SC. Again, 
there is possible time-reversal symmetry breaking state, 
although frustration in that case is relieved in a 
different (but related) manner.

\section{Acknowledgements}
We are grateful to V. Cvetkovi\' c for useful discussions.
Work at the Johns Hopkins-Princeton Institute
for Quantum Matter was supported by the U.\ S.\ Department of
Energy, Office of Basic Energy Sciences, Division of Materials Sciences and 
Engineering, under Award No.\ DE-FG02-08ER46544.

\bibliographystyle{apsrev}


\end {document}